# Population-specific design of de-immunized protein biotherapeutics

Benjamin Schubert[1,2,3,4*], Charlotta Schärfe[1,2,4], Pierre Dönnes[1,5], Thomas Hopf[3,4], Debora Marks[4], and Oliver Kohlbacher[1,2,6,7,8]

[1] Center for Bioinformatics, University of Tübingen, 72076 Tübingen, Germany

[2] Applied Bioinformatics, Dept. of Computer Science, 72076 Tübingen, Germany

[3] Department of Cell Biology, Harvard Medical School, Boston, 02115 Massachusetts, USA

[4] Department of Systems Biology, Harvard Medical School, Boston, 02115 Massachusetts, USA

[5] SciCross AB, 541 23 Skövde, Sweden

[7] Quantitative Biology Center, 72076 Tübingen, Germany

[8] Faculty of Medicine, University of Tübingen, 72076 Tübingen, Germany

[9] Biomolecular Interactions, Max Planck Institute for Developmental Biology, 72076 Tübingen, Germany

[*] Corresponding author:

E-mail: benjamin_schubert@hms.harvard.edu (BS)

# ABSTRACT

Immunogenicity is a major problem during the development of biotherapeutics since it can lead to rapid clearance of the drug and adverse reactions. The challenge for biotherapeutic design is therefore to identify mutants of the protein sequence that minimize immunogenicity in a target population whilst retaining pharmaceutical activity and protein function. Current approaches are moderately successful in designing sequences with reduced immunogenicity, but do not account for the varying frequencies of different human leucocyte antigen alleles in a specific population and in addition, since many designs are non-functional, require costly experimental post-screening.

Here we report a new method for de-immunization design using multi-objective combinatorial optimization that simultaneously optimizes the likelihood of a functional protein sequence at the same time as minimizing its immunogenicity tailored to a target population. We bypass the need for three-dimensional protein structure or molecular simulations to identify functional designs by automatically generating sequences using probabilistic models that have been used previously for mutation effect prediction and structure prediction. As proof-of-principle we designed sequences of the C2 domain of Factor VIII and tested them experimentally, resulting in a good correlation with the predicted immunogenicity of our model.

Protein-based drugs (biotherapeutics) are increasingly used to treat a wide variety of diseases[1,2]. Although biotherapeutics show high activity and specificity at the initiation of treatment, the gradual build-up of a patient immune response is a bottleneck for even wider usage[3]. This immune response involves the formation of anti-drug antibodies (ADAs) that target the biotherapeutic itself and cause loss of effect or adverse reactions[3-5]. A prominent example of this adverse effect is in the treatment of hemophilia A (HA) with coagulation Factor VIII, where ADAs develops in 10-15% of all HA patients and as much as 30% of those patients with the most severe form of HA[6]. Patients with the highest need for therapy are thus least likely to benefit. This correlation between severity of the disease and lack of efficacy follows from the fact that the immune system is more likely to recognize the therapeutic Factor VIII as foreign the more severe the natural mutation is, where mutations that cause a total loss of Factor VIII production are most strongly associated with ADA development[7,8].

The reduction of the immunogenicity has thus become a major step in a the development of a biotherapeutic[5]. The primary focus of reducing immunogenicity has been on humanized monoclonal antibodies (mAbs) that are comprised of foreign complementarity-determining regions in the variable regions, with the remainder of human origin, and, more recently, on fully human mAbs using bioengineering techniques[9,10]. However, these approaches are not generally applicable to other classes of biotherapeutics and even humanized and full human mAbs can still induce a clinically relevant anti-drug immune response, likely through the CD4$^+$ T-cell mediated adaptive immune system[11,12]. The CD4$^+$ T-cell activation is induced by the recognition of linear sequential peptides (called epitopes) derived from the therapeutic protein, which are presented on human leucocyte antigen (HLA) class II molecules of antigen presenting cells. Therefore, the systematic removal of these epitopes by sequence alteration (termed de-immunization) has been successfully used as an alternative approach to reduce the immunogenicity of mAbs and other therapeutic proteins[12-16].

In recent years, computational screening approaches have been developed to suggest protein sequences with reduced the overall immunogenicity. The simplest approaches focused solely on introducing point mutations to reduce the amount of CD4$^+$ T-cell epitopes by applying well-established epitope prediction methods[17-19]. However, the suggested mutations can have a significant impact on the stability and function of the protein. Naïve approaches not considering the structural impact on the protein will inevitably produce inactive designs. More advanced methods therefore try to exclude potentially harmful mutations by predicting their impact with various metrics[20]. The most recent approaches simultaneous optimize the number of deleted epitopes as well as the stability of the protein approximated either using structural or simple sequence information[21,22].

Recent advances in statistical protein modeling now allow to accurately infer the tertiary structure[23-27] and mutational effects[28-30] of proteins using evolutionary information contained in an multiple sequence alignment of a protein family. The statistical global pairwise entropy model used for protein inference accurately captures co-evolving sites within a protein which can be utilized to identify structural and functional important position using evolutionary couplings (EC) analysis, infer the protein structure, and predicted the effects of mutational changes (referred to as EVmutation).

In this work we present a novel formulation of the de-immunization problem that uses, for the first time to our knowledge, the maximum entropy model for protein design solving the inverse inference problem. Incorporating the maximum entropy model, as opposed to force-field based approaches such as FoldX[31], has four distinct advantages: (i) The statistical model does not require a known structure or depend on the conditions in which the structure was measured. (ii) It implicitly considers constraints on residues from interactions with ligands and other proteins, and (iii) models interactions between mutations rather than early filtering of deleterious singles. (iv) A de-immunization approach using the maxi-mum entropy model is likely to generate more viable structures as it minimizes potential damage to protein function at the same time as minimizing the immunogenicity of the biotherapeutic design. The latter can also be achieved by incorporating a force field, such as AMBER[32], into the optimization process [21,22], which however complicates the de-immunization formulation.

As the frequencies of HLA alleles differ drastically between populations, the immunogenicity of the biotherapeutic might differ as well. It is thus imperative to design a biotherapeutic for a specific target population considering their HLA allele frequencies, as opposed to treating each HLA allele equally important during the design process, as previous methods have done. We therefore developed a new quantitative immunogenicity objective that considers the HLA allele distributions within different populations. The resulting de-immunization model does not require known structural information about the protein, summarizes functional and structural information that might not be captured by a structure-based model, and considers the varying HLA frequencies in different populations. We also demonstrate how the resulting bi-objective combinatorial optimization problem can be formulated in a concise manner and solved efficiently for relevant problem sizes with a newly developed distributed solving strategy. An experimental validation of the resulting designs confirms that the algorithm can indeed lead to significantly reduced immunogenicity.

## RESULTS

### Formulation of the de-immunization problem

The problem of protein de-immunization can be described as identifying amino acid substitutions that reduce immunogenicity by removing T-cell epitopes while at the same time keeping the structure and function of the protein intact. We therefore define the problem of de-immunization as a bi-objective optimization problem. The first objective characterizes the immunogenicity of the target protein with re-spect to a set of HLA alleles by applying a start-of-the art, continuous immunogenicity prediction method (TEPITOPEpan[33]). The immunogenicity objective $I(S| H, P_H)$ combines the immunogenicity of each predicted epitope over a certain binding threshold weighted by the HLA allele frequencies $P_H$ of a specific target population represented by their prevalent HLA alleles $H^{34}$. The second objective $E(S)$ approximates the protein fitness via the statistical energy of the protein sequence, computed by the pairwise maximum entropy model inferred using a multiple sequence alignment (MSA) of the target protein family[23-27].

More formally, we define the protein de-immunization problem as follows: Given a protein sequence S of length n and a set $M_i$ of possible alterations per position $1 \leq i \leq n$. We seek a mutant S′ of S with k alterations for which $S'[i] \in M_i \ \forall \ 1 \leq i \leq n$ holds and that minimizes:

$$\text{argmin}_{S'} \ \left(I(S'|H,P_H), -E(S')\right)$$
$$\text{s.t.} \quad S' \in X,$$

The model therefore optimizes the trade off between these two objectives and produces a set of Pareto-optimal designs of the protein sequence.

As proof-of-principle, we tested the ability of the model to predict low immunogenic constructs of the C2 domain of Factor VIII as the domain is highly immunogenic and involved in the ADA development in hemophilia A patients when used therapeutically[35,36].

**Accurate predictions of mutation effects using EVmutation**

Evolutionary couplings computed from sequence alignments have been used successfully to predict the phenotypic effects of mutations[28-30,37], as well as the 3D structure shown in earlier work[23-27]. The ap-proach assigns an evolutionary statistical energy to any protein sequence that is hypothesized to correspond to the fitness of the molecule. The computation of the statistical energy of the protein and any changes to it after mutation is automatic and does not depend on computing or knowing the 3D structure. Therefore we reasoned that we could use this statistical model in a generative mode for design within the algorithmic de-immunization process.

As previous work on predicting the effect of mutations suggested that the model accuracy depends on the diversity of the sequence alignment and the ability to predict the 3D structure accurately[30], we used the precision of the total epistatic constraints between residue pairs as an approximation of the model validity. Overall, 70 long-range evolutionary coupled residue pairs (ECs) have a probability of at least 90% of being significant and 65 of these (93% precision) are close in space (less than 5Å; Figure 1A, Supplementary material S1) in a 3D structure of Factor VIII's C2 domain (pdb: 3hny[38]; Fig. 1B).

To assess how well EVmutation can predict the effects of specific mutations compared to force-field methods, we used EVmutation and FoldX predictions in a multinomial and logistic regression to predict hemophilia A severity (severe, moderate, and mild) based on patient data collected from the Factor VIII variant database (http://www.factorviii-db.org, Supplementary Table S2). Since the severity of HA is directly correlated with instability and malfunctioning of Factor VIII, the prediction of disease severity can be seen as a proxy for functional and structural effect prediction. The multinomial regression model, using the change in statistical energy between mutant and wild type as independent variable, shows a moderate ability to predict the clinical outcome (F1-micro of $0.65 \pm 0.09$, F1-macro of $0.47 \pm 0.07$).

# Factor VIII - C2 Domain

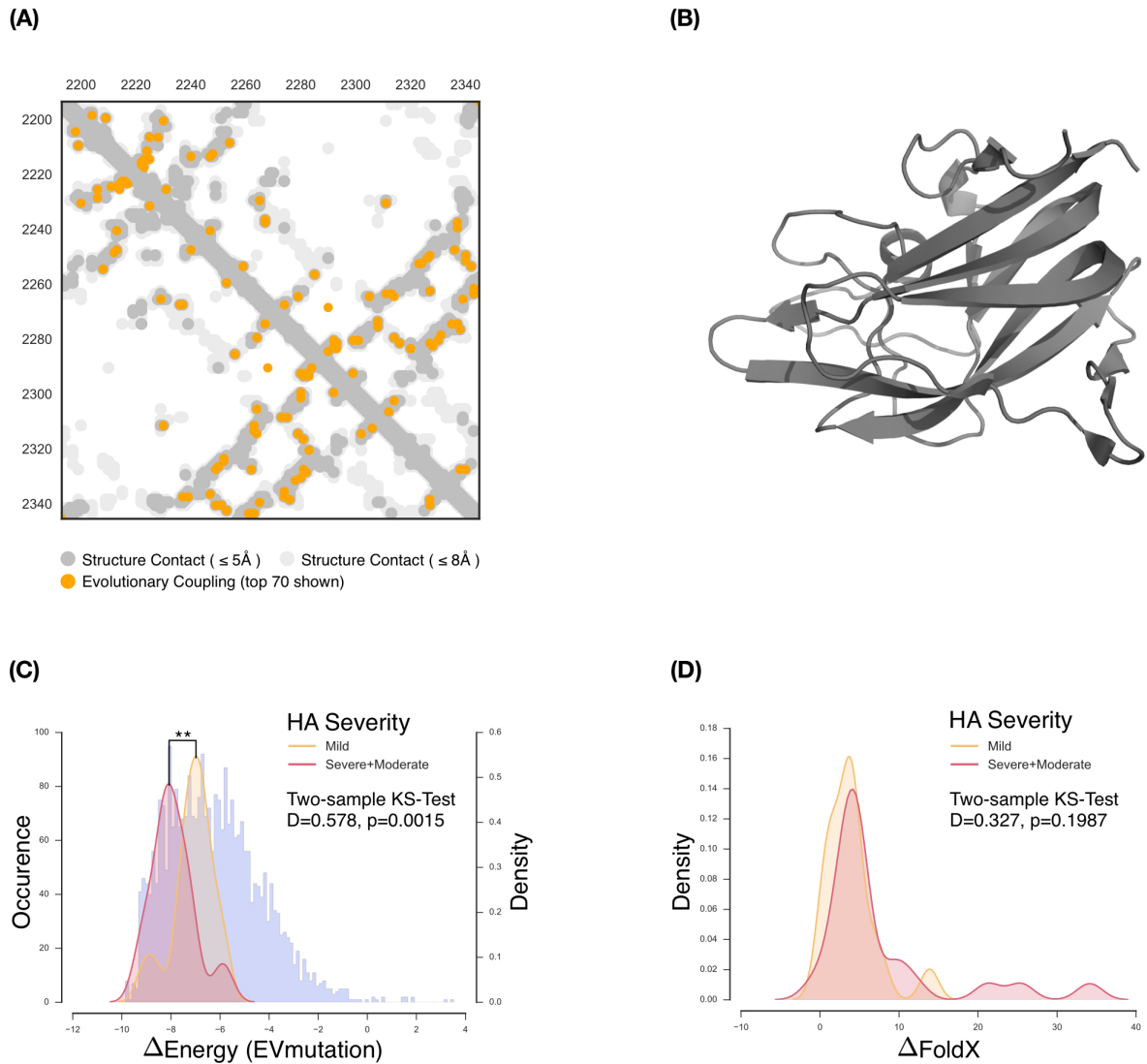

Figure 1: (A) Contact map of Factor VIII's C2 domain. The gray circles represent the known crystal structure (pdb: 3hny), while the orange dots represent predicted evolutionary couplings. (B) The tertiary structure of Factor VIII's C2 domain (pdb: 3hny). (C) Statistical energy change density separated by HA severity status (red and yellow) and the complete EVmutation landscape of single point mutations (blue). As expected, all known single point mutations with known HA severity status reside in the lower percentile of the energy landscape. The severe and moderate HA cases are clearly separable from the mild cases using EVmutation prediction. In contrast, the two distributions can note be clearly separated using FoldX predictions (D).

The performance of a FoldX-based multinomial regression model however was significantly worse (one-sided Wilcoxon signed rank test, V=13558, p-value < 2.2e-16; F1-micro of 0.49 ± 0.11, F1-macro of 0.35 ± 0.09). We combined the severe and moderate clinical classes and performed a logistic regression, which improved the prediction performance of the EVmutation-based model (weighted AUC of 0.72 ± 0.11, weighted F1-score of 0.73 ± 0.11; Supplementary Table S2; Fig. 1C). The FoldX-based logistic regression model was again outperformed by the EVmutation model (one-sided Wilcoxon signed rank test, V=15633, p-value < 2.2e-16; Fig 1D), but also yielded higher predictive power compared to its multinomial model (weighted AUC: 0.62 ± 0.11, weighted F1 0.58 ± 0.13).

**Strong immunogenic region contains functional sites.**

Initial prediction of epitopes using TEPITOPEpan[33] with the three most prevalent HLA alleles in the European population (accounting for 70% of the patients in Western Europe) resulted in 16 epitopes in 6 regions of the C2-domain of Factor VIII (UniProt: FA8_HUMAN). The region with the highest scoring immunogenicity (residues 2,312-2,340) had nine of the 16 predicted strong binding epitopes (Fig. 2A and 2B), making it a prime candidate region for de-immunization mutation design. However there is evidence that this very region might be of high functional importance for the protein; The region is enriched for conserved co-variation of residues and contains a known membrane-binding motif[38]. Eight of the top ten evolutionary couplings involve residues in the high immunogenic region (Fig. 2C). In general the region is enriched for strong evolutionary couplings (sign test, s = 124, n = 130, p-value < 2.2e-16, CI95 = [0.90, 0.98]; Fig. 2A). Hence there is a risk that mutations designed to minimize immunogenicity could be detrimental to protein function and the method we have developed here is specifically designed to minimize the risk of both.

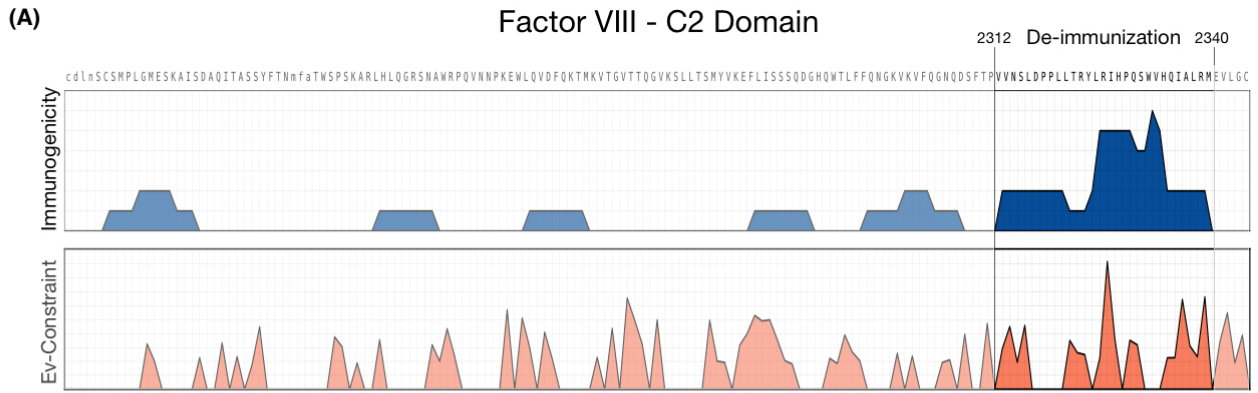

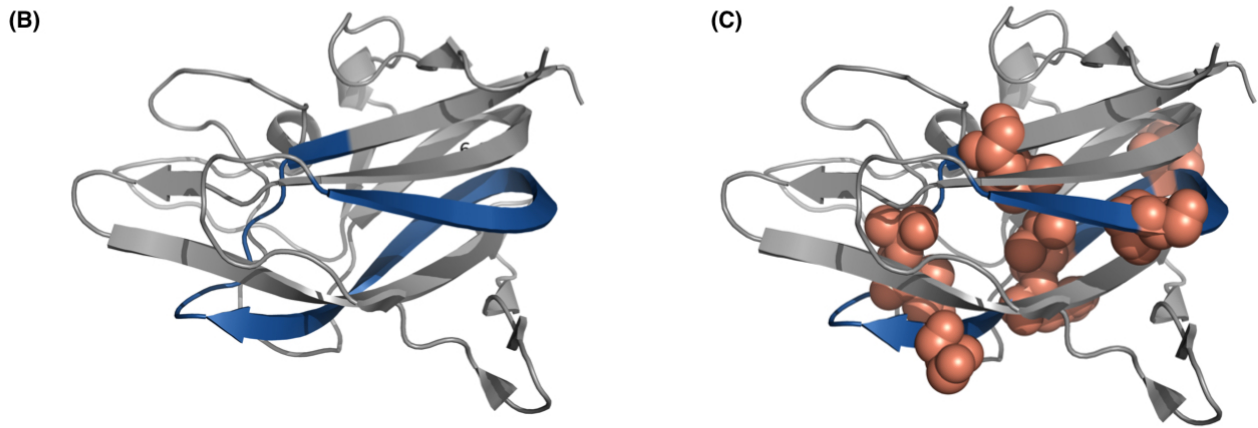

Figure 2: (A) Immunogenicity screening for three DRB1 alleles with TEPITOPEpan. The blue regions depict the cumulative immunogenicity scores per position and the orange regions depict the EC cumulative summary scores of top 70 ECs. Six immunogenic regions can be identified based on the in silico screening, where region 2,321 to 2,340 is the highest immunogenic region containing 9 out of 16 predicted epitopes and was thus chosen as de-immunization target. It is comprised of the highest evolutionary coupling pairs. (B) The tertiary structure of Factor VIII's C2 domain of with highlight immunogenic region selected for de-immunization redesign in blue. (C) The tertiary structure of Factor VIII's C2 domain with marked top eight EC (red spheres) that coincide with the identified immunogenic region (blue).

**De-immunization of Factor VIII's C2 domain**

We next used our bi-objective de-immunization model to design sequences of the identified highly immunogenic region resulting in 21 Pareto-optimal sequences with up to three simultaneous point mutations (Table 1, Fig. 3). Although the model was set up to constrain sequence substitutions solely to the identified immunogenic region, the resulting fitness change was optimized based on the interactions with all sites in that protein domain.

Table 1: De-immunization results for mutation loads of $k = 1,2,3$.

| ID | Mutation | Epitopes | ΔImmunogenicity | ΔEnergy | ΔFoldX |
|---|---|---|---|---|---|
| wt |  | 16 |  |  |  |
| 0 | V2333E | 11 | -0.38 | 1.14 | 0.95 |
| 1 | L2321F | 16 | 2.25 | 0.83 | 0.7 |
| 2 | Q2335H | 16 | 4.91 | 0.77 | 0.07 |
| 3 | Y2324L,V2333E | 9 | -2.16 | 6.47 | 0.56 |
| 4 | Y2324H,V2333E | 10 | -1.84 | 5.96 | 3.78 |
| 5 | R2326K,V2333E | 10 | -1.84 | 4.31 | 2.67 |
| 6 | L2321T,V2333E | 10 | -1.59 | 3.68 | 2.13 |
| 7 | L2321Y,V2333E | 11 | -1.01 | 3.48 | 1.7 |
| 8 | L2321F,V2333E | 12 | -0.99 | 1.97 | 1.66 |
| 9 | V2333E,Q2335H | 12 | 0.43 | 1.93 | 1 |
| 10 | L2321F,Q2335H | 17 | 4.3 | 1.47 | 0.57 |
| 11 | V2313M,Y2324L,V2333E | 8 | -2.52 | 7.99 | 0.16 |
| 12 | L2321T,I2327L,V2333E | 8 | -2.39 | 6.47 | 3.1 |
| 13 | L2321F,R2326K,V2333E | 10 | -2.21 | 5.32 | 3.36 |
| 14 | V2313M,L2321T,V2333E | 9 | -1.95 | 5.16 | 1.61 |
| 15 | L2321F,I2313V,V2333E | 10 | -1.92 | 4.92 | 3.29 |
| 16 | L2321F,I2313L,V2333E | 10 | -1.53 | 4.61 | 2.78 |
| 17 | V2313T,L2321F,V2333E | 10 | -1.36 | 4.58 | 1.49 |
| 18 | V2313M,L2321F,V2333E | 11 | -1.34 | 3.52 | 0.99 |
| 19 | L2321F,Y2324F,V2333E | 12 | -1.05 | 3.26 | 0.93 |
| 20 | L2321F,V2333E,Q2335H | 13 | -0.19 | 2.62 | 1.44 |
| 21 | L2321F,Y2324F,Q2335H | 17 | 4.24 | 2.55 | 0.21 |

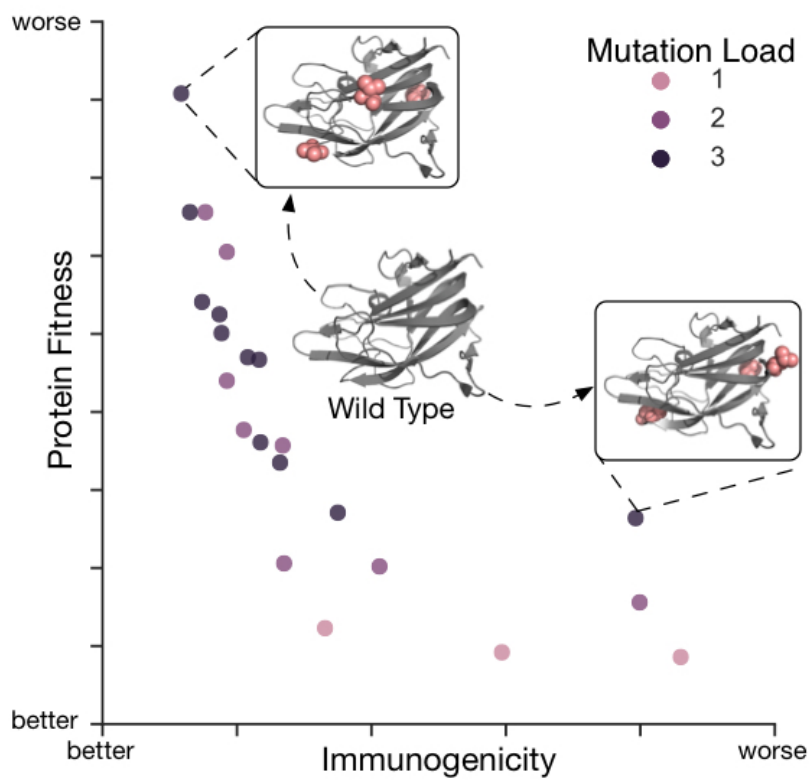

Figure 3: Pareto-front of de-immunized designs with k=1,2,3 mutations. Each design is a trade-off between the immunogenicity and the protein fitness function and represents a newly sequence (here represented as tertiary structures). The red spheres within the tertiary structures mark the mutated residues.

Even though none of the 21 designed sequences were predicted "fitter" than the wild–type, they were all close to the wild-type fitness. The computed fitness of 20 out of 21 designs resided in 95% percentile or higher when compared to the whole distribution of single, double, and triple mutations, suggesting that the protein would remain stable and functional. The sequence with the highest difference to wild-type fitness prediction (Design-11; V2313M, Y2324L, V2333E) was located in the 90% percentile and still close to WT fitness (reduction of 1.7%) and also had the maximal predicted reduction of immunogenicity (immunogenicity reduction of 45%) deleting eight out of nine epitopes of identified region. The next-best triple mutant (L2321T, I2327L, V2333E) resulted in the deletion of eight epitopes with an immunogenicity reduction of 42% and a fitness reduction of 1.28%.

Previous work that aims to increase the likelihood of a functional protein after mutation design, has used force-field based modeling, such as FoldX[31]. As to distinguish the differences between FoldX and the employed statistical sequence model we predicted the mutation effects of the 21 designs with EVmutation and FoldX (Table 1). EVmutation's predictions only moderately correlate to those using FoldX (r = 0.44, CI95 = [0.02, 0.73], t = 2.173, df = 20, p-value = 4.2e-2; Fig. 4). The two most deviating mutations between the two prediction methods were Design-11 (V2313M, Y2324L, V2333E) and Design-3 (Y2324L, V2333E), both of which introduced a mutation at a membrane-binding site[38]. FoldX predicted these designs comparatively less deleterious than EVmutation. One explanation for this discrepancy is that it would be harder for force-field based methods to capture the membrane binding constraints unless they were in the structure used.

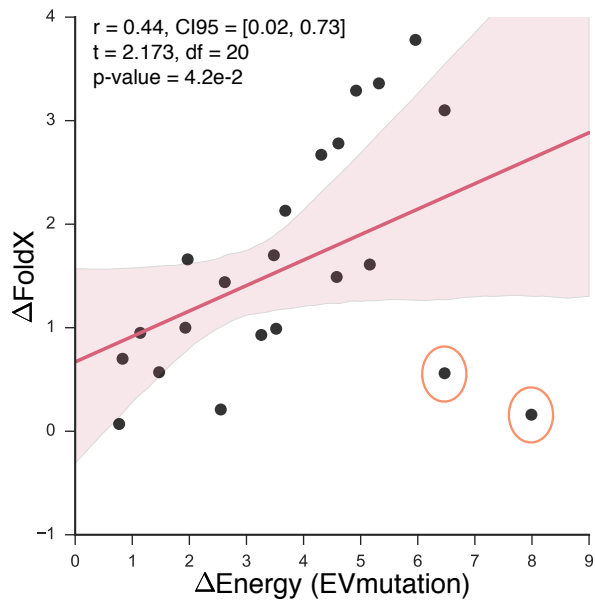

Figure 4: EVmutation and FoldX prediction correlations. The red line is a fitted linear regression, and the red tube represents its 95-confidence interval. The orange-circled dots are the two mutational designs with the highest discrepancy. FoldX predicted these two mutations less deleterious compared EVmutation, although both designs introduced a mutation at a membrane-binding site.

**Experimental validation**

To test the designs, we synthesized overlapping 15-mer peptides containing the introduced mutations and their wild-type counterparts. The peptides maximally covered the predicted epitopes around the mutations of all designed constructs that contained one and two mutations (Supplementary Table S2). The affinity of these peptides to the three HLA alleles was measured at time zero and after 24 hours (Methods, Supplementary Table S3 and S4). We linearly combined the measured affinity scores across alleles for each peptide respectively to produce a score that summarizes the overall immunogenicity (Fig. 5A).

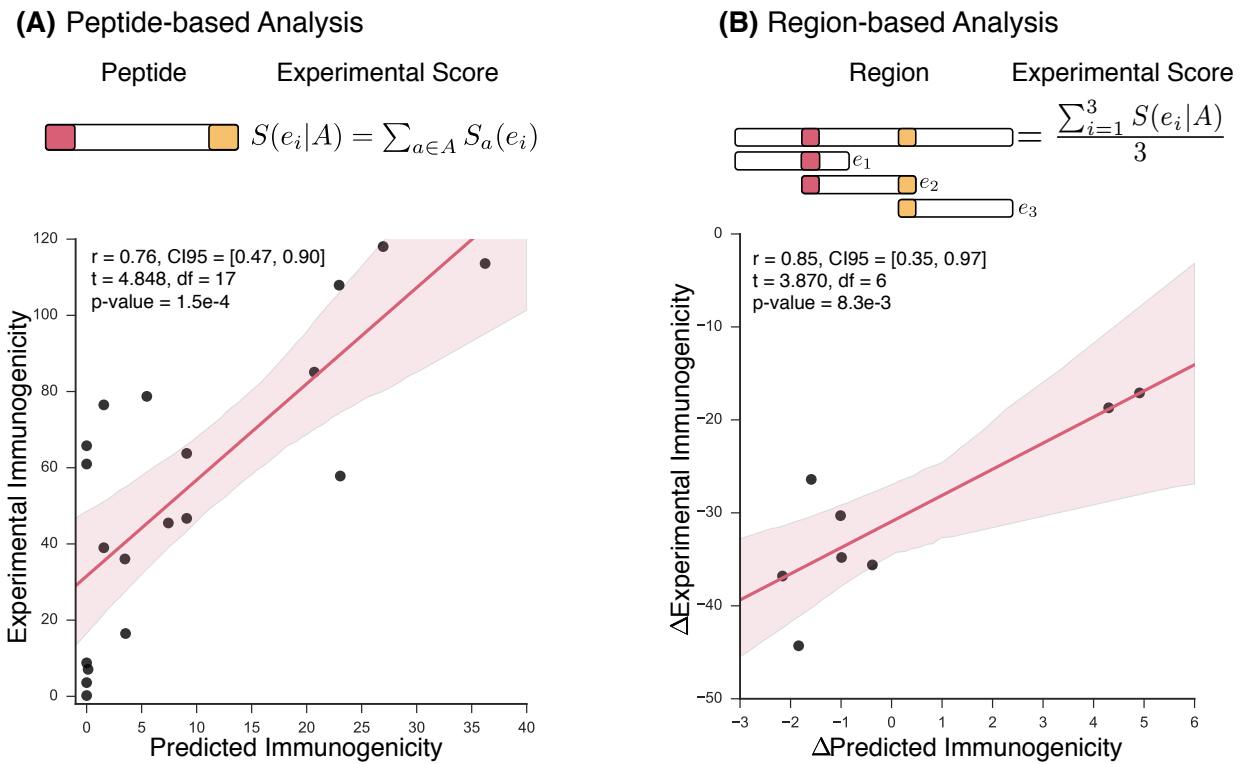

Figure 5: (A) Correlation of experimental and predicted immunogenicity of each peptide. The immunogenicity score of a peptide is defined as the linear combination of the individual experimental scores of each HLA allele $h \in H$. (B) Correlation of experimental and predicted immunogenicity of the whole redesigned region. The summarized immunogenicity score of the whole region is the linear combination of the overlapping peptides used to reconstruct the region, normalized by total number of peptides used. The red lines are a fitted linear regression and the red tubes represent their 95-confidence interval.

Overall, the measured and predicted immunogenicity of the tested peptides correlated well with r = 0.76 (CI95 = [0.47, 0.90], t = 4.848, df = 17, p-value = 1.5e-4; Fig. 5A). The highest reduction of 86.04% was observed for M-8 at while the highest increase in immunogenicity of 87.01% was observed for M-4 confirming the predicted decrease of 86.87% for M-8 and predicted increase of 86.37% for M-4.

Next we compared the predicted and measured gain or loss in immunogenicity for the whole region by reconstructing the targeted region using overlapping peptides. The measured scores were linearly combined (as previously) and then normalized to the number of overlapping peptides used in the reconstruction (Fig. 5B). The difference in the measured scores between the wild type and mutant regions can be thought of as a proxy for gain or loss in immunogenicity and correlate well to the predicted changes (r = 0.85 (CI95 = [0.35, 0.97], t = 3.870, df = 6, p-value = 8.3e-3; Fig. 5B). Measurements made after 24 hours were very similar (r = 0.94, CI95 = [0.86, 0.98], t = 11.81, df = 17, p-value = 1.3e-09).

**DISCUSSION**

This work introduced a novel method to reduce a protein's immunogenicity while maintaining its functionality requiring only sequence information of the target protein. The method uses a new immunogenicity objective that is, in contrast to previous approaches, accounting for relative epitope strength and integrates HLA allele frequency information of a target population. Since the HLA distribution can differ tremendously between populations this will influence the immunogenicity of a protein and hence T-cell epitopes that are prioritized during the de-immunization process.

The fact the highly immunogenic region, *a prior* identified during an *in silico* screening and independently described by others[19], coincides with a highly evolutionary connected as well as functionally important region underlines the need for methods that are capable of incorporating functional and structural integrity prediction in the de-immunization process. The de-immunization model introduced demonstrates the power of such approaches; the predicted immunogenicity of the complete domain could be reduced by 45% without disrupting the fitness landscape extensively. Moreover, the observed highly significant correlations between measured and predicted immunogenicity both on individual peptide and (reconstructed) segment level affirmed that the underlying assumptions made by the model are sufficient to predict the influence of mutation in terms of immunogenicity.

In the particular case of this Factor VII domain, we found no advantage to structure and force-field base approaches to assessing the effect of clinical mutation classification; structure-based approaches may even be a disadvantage when structure information is incomplete (e.g., binding partners not present). This suggests that sequence information may be sufficient for de-immunization design, and is consistent with the previous observation that sequences alignments can be used to identify constrained interacting residues across biomolecules as well as the effect of mutations[30,39]. However, high-quality diverse sequence alignments are not always available.

In summary, we proposed a novel de-immunization model that integrates quantitative immunogenicity optimization with sequence-based fitness optimization and used the approach to design novel C2 domains of Factor VIII that can be further validated for clinical application using mouse models or T-cell proliferation assays based on PBMCs of HA patients. The approach will allow bioengineers to reliably explore the design space of the target protein to select promising candidates for experimental evaluation.

**METHODS**

**Derivation of the immunogenicity objective**

The first objective of the de-immunization model is an adaptation of the immunogenicity score introduced by Toussaint *et al.* for epitope selection in the context of *in silico* vaccine design and is defined as follows [34]:

$$I(E|H, P_H) = \sum_{e \in A} \sum_{h \in H} p_h \cdot i_{e,h},$$

with $A$ being a set of epitopes, $i_{e,h}$ the immunogenicity of epitope $e \in A$ bound to HLA allele $h \in H$. It assumes that each epitope independently influences the immune response with respect to all considered HLA alleles. The contribution of an HLA allele $h \in H$ is directly proportional to its probability $p_h$ of occurring within the target population $H$.

**Derivation of the protein fitness objective**

The second objective is a evolutionary statistical energy of sequences computed by a pairwise maximum entropy model of protein families. Under these family-specific models, the probability for a protein sequence $(X_1, ..., X_n)$ of length $n$ is defined as

$$P(X_1, \ldots X_n) = \frac{1}{Z} e^{-E(x)}$$

$$E(X) = \sum_{1 \le i \le j \le n} J_{ij}(X_i, X_j) + \sum_{1 \le i \le n} h_i(X_i)$$

where the pair coupling parameters $J_{ij}(X_i, X_j)$ describe evolutionary co-constraints on the amino acid configuration of residue pairs $i$ and $j$ for all amino acids and the parameters $h_i(X_i)$ corresponds to single-site amino acid constraints. The partition function $Z$ is a global normalization factor summing over all possible amino acid sequences of length $n$[23,24]. The parameters $J_{ij}$ and $h_i$ are inferred from a protein family sequence alignment using an iterative approximate maximum likelihood inference scheme (pseudo-likelihood maximization) under $l_2$-regularization to prevent overfitting. Given an inferred probability model for a family, the statistical energy $-E(X)$ can be used to quantify the fitness of specific sequences. Recent work has demonstrated that changes of $E(X)$ quantitatively correspond to the experimental phenotypic consequences of mutations, including effects on protein stability and organismal growth[30]. In order to maintain protein function while minimizing immunogenicity, the second objective function is defined as the minimization of $-E(X)$.

Evolutionary coupling (EC) strength between pairs of positions *i* and *j* is computed using the Frobenius norms of the matrices $J_{ij}$ with subsequent correction for finite sampling and phylogenetic effects (average product correction)[26]. The evolutionary couplings are predictive of residue proximity in many protein families, and the cumulative score of one position to all others (EC enrichment score) is indicative of functionally and structurally important positions [26].

**Derivation of the integer linear program representation of the de-immunization model**

We solve the stated de-immunization problem as a bi-objective mixed integer linear program (BOMILP). Solving a BOMILP finds all Pareto-optimal solutions to linear objectives with affine constraints and additional integrality constraints on a subset of the variables. The model is based on Kingsford *et al.*'s ILP formulation of the side-chain placement problem [40]. But instead of selecting energetically favorable rotamers, we encode each state of the model as a possible amino acid substitution at each position. A binary decision variable $x_{i,a}$ for each position $i \in \{1..n\}$ and each possible variation $a \in M_i$ is introduced with $x_{i,a} = 1$ if this variant will be part of the final mutant $S'$. An additional binary variable is introduced for each pair of variants and positions notated $w_{i,j,a,b}$ with $w_{i,j,a,b} = 1$ if variant $a$ at position $i$ and variant $b$ at position $j$ have been selected as part of the solution $S'$. These variables are associated with their inferred fitness terms $h_{i,a}$ and $J_{i,j,a,b}$ to form the second objective function (Table 2 O2).

The immunogenicity objective, in contrast to the problem formulation of Toussaint *et al.,* in which the immunogenicity of each candidate epitope $e \in A$ could be pre-calculated, does not have an easy ILP representation. It has to directly incorporate prediction methods to approximate the immunogenicity of the current mutant $S'$. Therefore, we use TEPITOPEpan[33] as internal prediction engine since it has been demonstrated to have good predictive power and can be easily integrated into the ILP framework. More advanced, machine learning-based prediction models cannot readily be integrated into the problem formulation. With linear (matrix-based) methods, the integration is possible by scoring each peptide generated with a sliding window of width $e_n$ for each allele $h \in H$ independently by summing over TEPITOPEpan's position specific scoring matrix $\Phi(h, a, j) \rightarrow \mathbb{R}_{\geq 0}$ for amino acid $a \in M_{i+j}$ at position $i + j$ with $i \in \{1..(n-e_n)\}$ and $j \in \{0..(e_n - 1)\}$. To only consider predicted binding epitopes, the binding threshold $\tau_h$ of each HLA allele $h \in H$ is subtracted from the sum score of an epitope and the maximum of 0 and the cumulative score is taken. The sum of all epitope scores for a particular allele $h \in H$ is than weighted by its population probability $p_h$. To make the prediction scores comparable across HLA alleles, the position specific scoring matrices of TEPITOPEpan were z-score normalized and the binding thresholds adjusted accordingly. The final immunogenicity score consists of the sum of all allele-wise weighted sums (Table 2, O1).

Table 2: Bi-objective integer linear program formulation of the de-immunization problem.

**Definition:**

**Objectives:**

(O1) $\min_{x,w} \quad \sum_{h \in H} p_h \cdot \sum_{i=1}^{n-e_n} \max\left(0, \left(\sum_{j=0}^{e_n-1} \sum_{a \in M_{i+j}} x_{i,a} \cdot \phi(h, a, j)\right) - \tau_m\right)$

(O2) $\min_{x,w} \quad \sum_{i=1}^{n} x_{i,a} \cdot h_{i,a} + \sum_{i=1}^{n} \sum_{1 \leq i < j \leq n} w_{i,j,a,b} \cdot J_{i,j,a,b}$

**Constraints:**

(C1) $\sum_{a \in M_i} x_{i,a} = 1$ $\quad\quad\quad \forall i \in \{1..n\}$

(C2) $\sum_{b \in M_j} w_{i,j,a,b} = x_{i,a}$ $\quad\quad\quad \forall i, a \in M_i, i > j \in \{1..n\}$

(C3) $\sum_{a \in M_i} w_{i,j,a,b} = x_{j,b}$ $\quad\quad\quad \forall j, b \in M_i, j < i \in \{1..n\}$

(C4) $\sum_{i=1}^{n} \sum_{a \in W_i} (1 - x_{i,a}) \leq k$

To construct a consistent model, three constraints have to be introduced guaranteeing that only one amino acid per position is selected (Table 2, C1) and that only pairwise interactions are considered for selected variants (i.e., $J_{i,j,a,b} = 1 \leftrightarrow x_{i,a} = 1 \bigwedge x_{j,b} = 1$, see Table 2 C2 and C3). Constraints C2 and C3 can be further relaxed by dividing the pairwise fitness values into positive and negative sets [40], which is done in practice but disregarded here for ease of presentation. To be able to restrict the mutant to a specific number of introduced variations, constraint C4 limits the number of deviating amino acids to the wild type sequence $W$. A detailed formulation of the complete optimization problem can be found in Supplementary Table S6.

**Pre-processing**

To reduce the search space, a filtering approach based on position specific amino acid appearance frequency $f_i(a)$, i.e. conservation can be applied. Only amino acids at position $i \in \{1..n\}$ exceeding a certain frequency threshold $\zeta$ are considered as possible substitution at a site. Hence, the set of possible substitutions per position is defined as $M_i := \{a \in \Sigma \mid f_i(a) \geq \zeta\}$. The wild type amino acid is additionally added if it does not exceed the frequency threshold. This filtering is based on the assumption that variants that are not or infrequently observed are harmful due to either destabilizing effects, reduction of function, or intervening effects with interaction partners.

**Solving a bi-objective integer program**

Special strategies have to be applied to solve a BOMILP. Popular methods to solve discrete multi-objective problems include the $\varepsilon$-constraint[41], perpendicular search[42], and the augmented weighted Chebychev method[43]. All have their own limitations. The recently published rectangle splitting approach tries to overcome these[44,45]. For solving the de-immunization problem, we developed a parallel two-phase version of the rectangle-splitting approach that can exploit the parallel nature of the algorithm and can effectively utilize modern distributed computing resources. In the following we sketch the newly developed two-phase approach.

First we introduce necessary notations and concepts (adopted from Boland *et. al.*). Let $z^1 = (z_1^1, z_2^1)$ and $z^2 = (z_1^2, z_2^2)$ be two points in solution space with $z_1^1 \leq z_1^2$ and $z_2^2 \leq z_2^1$. Further we define $R(z^1, z^2)$ to be the rectangle spanned by $z^1$ and $z^2$. A nondominated point within $R(z^1, z^2)$ can be found with the following sequential operation (see proof in[44]):

$$
\begin{aligned}
(1) \quad & \overline{z_1} = \min_{x \in \chi} z_1(x) \\
s.t: \quad & z(x) \in R(z^1, z^2) \\
(2) \quad & \overline{z_2} = \min_{x \in \chi} z_2(x) \\
s.t: \quad & z(x) \in R(z^1, z^2) \text{ and } z_1 \leq \overline{z_1}
\end{aligned}
$$

These operations will be denoted as $\tilde{z} = \text{lex} \min_{x \in X} \{z_1(x), z_2(x): z(x) \in R(z^1, z^2)\}$.

As a first step of the two-phase parallel rectangle-splitting approach the boundaries of the Pareto front are calculated by solving

$$z^T = \text{lex} \min_{x \in X} \{z_1(x), z_2(x): z(x) \in R((-\infty, \infty), (-\infty, \infty))\}$$

and

$$z^B = \text{lex} \min_{x \in X} \{z_2(x), z_1(x): z(x) \in R((-\infty, \infty), (-\infty, \infty))\}$$

in parallel (Fig. 6A). Then, the search space within $R(z^T, z^B)$ is evenly constraint based on boundary conditions enforced w.l.o.g. on $z_1$ (Fig. 6B). The boundaries are calculated for a predefined number of constraints *m* with:

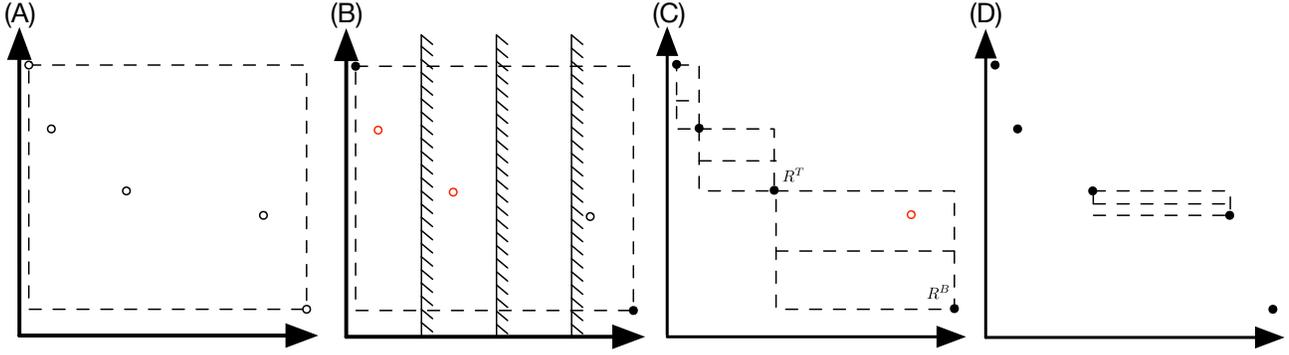

Figure 6: Depiction of the parallel two-phase rectangle splitting approach. (A) First, the boundaries of the Pareto front are identified. (B) Then, the space between the boundaries is evenly divided and searched in parallel for nondominated points using the ε-constraint method. (C) The identified nondominated points are used to initiate rectangle search spaces which can be processed in parallel using the standard rectangle-splitting approach, by splitting the rectangle in half and searching independently the bottom and top half (D). If the corner points of the rectangles are found during the search, it is proofs, that no further nodominated point resides within the search space and all points have been identified.

$$\tau_i^{z_1} = z_1^T + \frac{i \bullet (z_1^B - z_1^T)}{m} \;\; with \; 1 \leq i \leq m.$$

Each section of the separated search space can be independently searched by solving $z^i =$ lex $\min_{x \in X}\{z_1(x), z_2(x): z(x) \in R((\tau_i^{z_1}, z_2^T), z^B)\}$ and the resulting new nondominated points can be used as initial approximation of the Pareto front. The found nondominated points might contain duplicates and also might not resemble the complete Pareto front. Therefore, it is necessary to perform a refinement of the Pareto front to find the remaining nondominated points. To this end, the nondominated points are sorted in nondecreasing order such that $z_1^1 \leq z_1^2 \leq \cdots \leq z_1^k$. Each consecutive pair of points spans a search rectangle $R(z^i, z^j)$ with $i \leq j$. These rectangles can now be searched in parallel by the rectangle-splitting algorithm (Fig. 6C). The search rectangles are split in half. First, the bottom half $R^B$ is searched by solv-

ing:

$$\bar{z}^1 = \text{lex} \min_{x \in X} \left\{ z_1(x), z_2(x) \colon z(x) \in R\left( \left( z_1^i, \frac{z_2^i + z_2^j}{2} \right), z^j \right) \right\}$$

If a nondominated point is found, the upper half $R^T$ is further restricted and spans now $R\left( z^i, \left( \bar{z}_1^i - \epsilon, \frac{z_2^i + z_2^j}{2} \right) \right)$ in which $\bar{z}^2 = \text{lex} \min_{x \in X} \left\{ z_2(x), z_1(x) \colon z(x) \in R\left( z^i, \left( \bar{z}_1^i - \epsilon, \frac{z_2^i + z_2^j}{2} \right) \right) \right\}$ is searched. Each newly found point spans a new independent search rectangle $R(z^i, \bar{z}^2)$ and $R(\bar{z}^1, z^j)$ with its adjacent point. These rectangles are searched in parallel with the described procedure (Fig. 6D). If the search operation yielded the known point $z^j$ for $R\left( \left( z_1^i, \frac{z_2^i + z_2^j}{2} \right), z^j \right)$ and $z^i$ for $R\left( z^i, \left( \bar{z}_1^i - \epsilon, \frac{z_2^i + z_2^j}{2} \right) \right)$ accordingly proofs that the area does not contain further nondominated points. The search procedure is carried out until the complete search space has been explored.

**Inference of the maximum entropy sequence model**

Multiple sequence alignments (MSA), created by JackHMMER[46], were used for the inference of the maximum entropy models of the Factor VIII C2-domain (residues 2,188-2,345 of FA8_HUMAN, Supplementary material S1). To optimize residue coverage and MSA diversity, the alignment was created using five search iterations at an E-value threshold of $10^{-20}$. Sequences with 70% or more gaps and columns with over 50% gaps were excluded from subsequent statistical inference. To reduce the influence of sampling bias in the inference step, sequences were clustered at a 90% identity threshold (theta 0.9), and reweighted by the inverse of the number of cluster members. The parameters of the pairwise maximum entropy model and evolutionary couplings were then inferred using EVfold[PLM][47] with pseudo-likelihood maximization[48] (Supplementary material S1). Substitution effects were derived by calculating the difference between the wild-type and the mutant statistical energy[30]. The validity of the maximum entropy model was verified by using the precision of the inferred top evolutionary couplings (ECs) between residue pairs compared to an existing 3D structure. To identify the top ECs a Gaussian-lognormal mixture model was inferred based on the overall score distribution[27] and the ECs within the tail of the distribution (ECs with a probability ≥ 0.90 of belonging to the lognormal) were used for model quality assessment[27].

**Hemophilia a severity data and regression analysis**

Single point mutation data with known patient severity status was extracted from the Factor VIII variant database (http://www.factorviii-db.org). The data was filtered for mutations residing within the C2 domain, which resulted in 40 data points in total (Supplementary Table S5). The severity status of each patient was determined based on a one-stage Factor VIII:C and categorized into three classes – severe (<1%), moderate (1-5%), and mild (>5%). The data points were unevenly distributed across the classes with 15 severe, 8 moderate, and 17 mild cases.

To train and validate the multinomial and logistic regression models, the data was randomly divided into training and test set (70:30%-split) in a stratified manner. This process was repeated two hundred times and the prediction performance averaged over the runs.

**Experimental design**

In order to experimentally verify our *in silico* predictions for Factor VIII, we utilized the commercial REVEAL HLA-Peptide binding assay of ProImmune (www.proimmune.com). In addition to binding, stabilities of HLA-peptide complexes were also investigated. Peptides were synthesized using the PEP screen custom library synthesis method, yielding high purity peptides for experimental analysis. HLA-peptide binding was assessed for the three HLA-DRB1 alleles used in this study (DRB1*1501,

DRB1*0301 and DRB1*0701). In short, the method compares the affinity of the studied peptide to the affinity of a high-affinity control peptide. Each peptide is then scored for binding to a certain HLA molecule relative to the score of the control peptide, and reported as the percentage of the signal generated by the control peptide. In order to assess the stability of binding, the amount of bound peptide is measured at time zero and after 24 hours. From these measurements a stability index is calculated.

**Implementation**

The two-phase rectangle-splitting solver was implemented in Python 2.7 using the CPLEX package, Numpy 1.4, and Polygon 2.0.6 package. CPLEX 12.6 was used as backend to solve the BOMILP models.

Structure-based fitness prediction for validation purposes of the de-immunized Factor VIII C2 domain constructs were performed with FoldX[31] using default settings for the obtained mutations. TEPITOPEpan 1.0 was used for epitope prediction.

**Statistical analysis**

The multinomial and logistic models were fit and evaluated using Scikit-learn 0.18[49]. The statistical analysis was conducted with R 3.0.2. Statistical significance was considered at $\alpha = 0.05$. The specific statistical tests used are indicated in the figures or in the results section.

**Code availability**

The solver implementation, the ILP model in AMPL format, and the instances solved are available at https://github.com/FRED-2/EVdeimmunization released under a 3-claus BSD license.

## ACKNOWLEDGEMENTS

*Author's contributions*: BS developed the mathematical model and bi-objective solver. BS and PD designed the experiment. BS, CS, PD analyzed the data. TH contributed source code. TH and DM helped analyzing the data. BS, CS, PD, TH, DB, and OK wrote the paper. OK designed the study.

*Funding*: This study was partially funded by Deutsche Forschungsgemeinschaft (SFB 685/B1, KO 2313/6-1) and Bundesministerium für Bildung und Forschung (01GU1106).

*The authors declare no conflict of interest*.

# ADDITIONAL INFORMATION

**Material S1:** The long-range evolutionary couplings of Factor VII C2 domain, and the multiple sequence alignments used for model inference.

**Table S2**: Known Factor VIII C2 domain single point mutation with severity status, and predicted statistical energy change with EVmutation and FoldX.

**Table S3**: Synthesized peptides sequences.

**Table S4**: Experimental immunogenicity score measured per peptide and HLA allele.

**Table S5**: Summarized experimental immunogenicity scores of the reconstructed immunogenic region.

**Material S6:** Detailed description of the de-immunization BOMILP formulation

# ABBRIVIATIONS

| | |
|---|---|
| ADA | Anti-drug antibody |
| BOMILP | Bi-objective mixed integer linear program |
| CD | Cluster of differentiation |
| EC | Evolutionary Couplings |
| HA | Hemophilia A |
| HLA | Human leucocyte antigen |
| ILP | Integer linear programming |
| mAb | Monoclonal antibody |
| MSA | Multiple sequence alignment |
| PBMC | Peripheral blood mononuclear cell |